\documentclass[a4paper,11pt]{article}
\usepackage{pos}

\newcommand\bef{\begin{figure}}
\newcommand\eef[1]{\label{fg:#1}\end{figure}}
\newcommand\bec{\begin{center}}
\newcommand\eec{\end{center}}
\newcommand\besf{\begin{subfigure}}
\newcommand\eesf[1]{\label{sfg:#1}\end{subfigure}}
\newcommand\beq{\begin{equation}}
\newcommand\eeq[1]{\label{#1}\end{equation}}
\newcommand\beqa{\begin{eqnarray}}
\newcommand\eeqa[1]{\label{#1}\end{eqnarray}}
\newcommand\bet{\begin{table}}
\newcommand\eet[1]{\label{tb:#1}\end{table}}
\newcommand\best{\begin{subtable}}
\newcommand\eest[1]{\label{stb:#1}\end{subtable}}
\newcommand\betb{\begin{center}\begin{tabular}}
\newcommand\eetb{\end{tabular}\end{center}}
\newcommand\beit{\begin{itemize}}
\newcommand\eeit{\end{itemize}}

\definecolor{DarkGreen}{rgb}{0.00,0.29,0.00}
\definecolor{DarkRed}{rgb}{0.79,0.00,0.00}

\def\prsp#1#2%
  {\mathop{}%
   \mathopen{\vphantom{#2}}^{#1}%
   \kern-\scriptspace%
   #2}

\author*[a]{M. Padmanath}

\affiliation[a]{Helmholtz Institut Mainz, Staudingerweg 18, 55128 Mainz, Germany \\
GSI Helmholtzzentrum für Schwerionenforschung, Darmstadt (Germany)}

\emailAdd{pmadanag@uni-mainz.de}
\emailAdd{papppan@gmail.com}

\title{Charm (and bottom) baryons and charmonium excitations from the lattice\footnote{MITP/21-040}}
\ShortTitle{Heavy baryon and charmonium spectrum from Lattice QCD}

\abstract{
This report discusses some recent investigations of the heavy hadron spectra using 
lattice QCD. The first half addresses multiple precision determinations of the masses 
of charm (and bottom) baryons. Recent lattice results in the tetraquark and the dibaryon
sectors are also presented. The second half focuses on new exploratory studies of 
the excited charmonium spectra in the vector and scalar channels. Along the way, lattice 
results are compared with the experimental results, wherever they are available.
}

\FullConference{%
  *** 10th International Workshop on Charm Physics (CHARM2020), ***\\
  *** 31 May - 4 June, 2021 ***\\
  *** Mexico City, Mexico - Online ***
}

\begin{document}
\maketitle

\section{Introduction}\label{Intro}

The past decade has been splendid for heavy hadrons with several discoveries both 
in conventional and exotic channels. Among these are the doubly charmed baryon 
\cite{Aaij:2017ueg}, the charmonium-nucleon pentaquark resonances \cite{Aaij:2015tga}, 
five narrow $\Omega_c$ resonances \cite{Aaij:2017nav}, and a handful of heavy hadrons 
with four quark content. Detailed accounts of various discoveries and the 
hadron properties can be found in recent reviews of this subject \cite{Esposito:2016noz,Olsen:2017bmm}. The 
expanding family of discovered heavy hadrons motivates theoretical investigations of 
such and similar excitations and their properties, which can be tested in experiments.

Remarkable progress in lattice QCD, which is an {\it ab-initio} non-perturbative 
approach that can be systematically improved to achieve complete control over all 
uncertainties, has allowed modern investigations to make precision measurements 
in QCD that are of phenomenological interest. This advancement is reflected in 
several lattice publications on different aspects of QCD, including the 
hadron spectrum determination in recent years (see Lattice yearly conference 
proceedings {e.g.} Ref. \cite{Proceedings:2020rxe}). Most of the investigations 
are made in the isospin symmetric limit, where masses of up and down quarks are chosen 
to be the same, and with a heavier than physical light quark mass. More recently, 
studies are also performed directly at the physical values for several heavy flavor 
physics observables as presented in the FLAG review\cite{FlavourLatticeAveragingGroup:2019iem}. 
While many calculations utilize lattice QCD gauge configurations that account 
for the dynamics of light and strange quarks during the simulation ($N_f=2+1$), 
some newer studies use ensembles that also incorporate the charm quark dynamics 
during the gauge field generation ($N_f=2+1+1$). Significant efforts have also  
been put to include the strong and QED isospin breaking effects using lattice 
techniques, but for heavy hadrons such studies are limited \cite{Borsanyi:2014jba}.

Hadron spectroscopy using lattice techniques commonly proceeds through the extraction 
of finite-volume energy spectrum ($E_n$) on a discretized space-time. This is achieved by 
computing two-point correlation functions 
\begin{equation}
C_{ij}(t_f-t')=\langle O_{i}(t_f)O^{\dagger}_{j}(t') \rangle=\sum_{n} \frac{Z_i^{n}Z_j^{n*}}{2 E_n}e^{-E_n(t_f-t')},
  \label{eq:2-1}
\end{equation}
where $O_{i}(t)$ is the hadronic current with the desired quantum numbers and 
$Z_i^{n} = \langle O_{i}|n\rangle$ is the operator state overlap. $O_{i}(t)$ couples 
to all the states including single-particle levels, and their radial excitations,
as well as multi-particle levels with these quantum numbers. A standard practice to 
extract the energy spectrum is to compute matrices of correlation functions between 
a basis of interpolators with the same quantum numbers \cite{Basak:2005ir,Dudek:2010wm} and to solve the Generalized EigenValue Problem (GEVP)
\cite{Michael:1985ne} 
\begin{equation}
C_{ij}(t)v_j^n(t-t_0)=\lambda^n(t-t_0)C_{ij}(t_0)v_j^n(t-t_0).
  \label{eq:2-2}
\end{equation}
$E_n$s are extracted from the large $t$ behavior of the eigenvalue correlators 
$\lambda^n(t-t_0)$.

We classify the modern-day lattice calculations for hadron spectroscopy into 
two categories. In the first category, one computes the energy spectrum on 
the lattice and attributes the finite-volume energy levels to the strong 
interaction stable hadrons. The mass estimates are extracted directly from 
simple fits to the large $t$ behavior of the two-point correlation functions. 
A handful of ground-state baryons are stable with respect to strong decays, and their 
masses can be extracted precisely on the lattice. Multiple lattice groups 
have been performing detailed systematic investigations of stable hadrons. 
A summary and details of various such studies can be found in the reviews 
\cite{Liu:2016kbb,Padmanath:2018zqw}. More complicated systems (such as 
hadrons close to the elastic threshold, tetraquark, and pentaquark systems 
that have received attention recently) are also treated on a similar 
footing as stable hadrons in several recent exploratory investigations 
on the lattice (see Section \ref{beyond}).

The trivial connection of the finite-volume energy levels with the 
infinite-volume spectrum is no longer valid for hadrons close to and above 
the elastic threshold, which falls into the second category. A standard 
procedure to extract the infinite-volume two-particle scattering amplitudes 
from finite-volume energy spectrum is through the quantization condition, 
\begin{equation}
det(K^{-1}-B) = 0,
  \label{eq:2-3}
\end{equation}
first derived by L\"uscher for elastic scattering of two spinless particles 
in the rest frame\cite{Luscher:1986pf}. Here $K$ is related to the 
infinite-volume scattering amplitudes and $B$ is built out of known 
mathematical functions of the system's total energy $E_{cm}$. Given the 
finite-volume energy levels and a parametrization of $K$, one looks for 
the best-fit parameters that satisfy Eq. \ref{eq:2-3}. The infinite-volume 
scattering amplitudes built from these best-fit parameters are then 
investigated across the complex s-plane for pole singularities 
related to discovered hadrons. Investigating the model independence of 
the pole positions in the extracted scattering amplitudes using different 
parametrizations of $K$ further demonstrates the robustness of the 
lattice determinations \cite{Dudek:2014qha}. A detailed review 
on methodologies for treating the hadronic resonances on the lattice and 
various lattice calculations along these lines can be found in Ref. 
\cite{Briceno:2017max}. A summary of results from more recent calculations 
in the light hadron and heavy meson sector are discussed in the proceedings \cite{Edwards:2020rbo}.
Recent theoretical and numerical advancements in treating three-particle scattering 
in the finite-volume can be found in Refs. \cite{Hansen:2019nir,Mai:2021lwb,Hoerz:2021zqw}.

This report presents results from some recent lattice determinations 
of the spectra of heavy hadrons. Section \ref{cbbaryons} focusses on heavy 
baryons, whereas section \ref{beyond} on heavy four-quark and six-quark systems. 
In section \ref{mesres}, we discuss a recent investigation of excited and 
exotic charmonium resonances and  summarize in Section \ref{summar}.

%%%%%%%%%%%%%%%%%%%%%%%%%%%%%%%%%%%%%%%%%%%%%%%%%%%%%%%%%%%%%%%%%%%%%%%%%%%%%%%%%%%%%%%%%%
%%%%%%%%%%%%%%%%%%%%%%%%%%%%%%%%%%%%%%%%%%%%%%%%%%%%%%%%%%%%%%%%%%%%%%%%%%%%%%%%%%%%%%%%%%

\section{Charm and bottom heavy baryons}\label{cbbaryons}

\begin{figure}
\includegraphics[height=3.8cm,width=6.0cm]{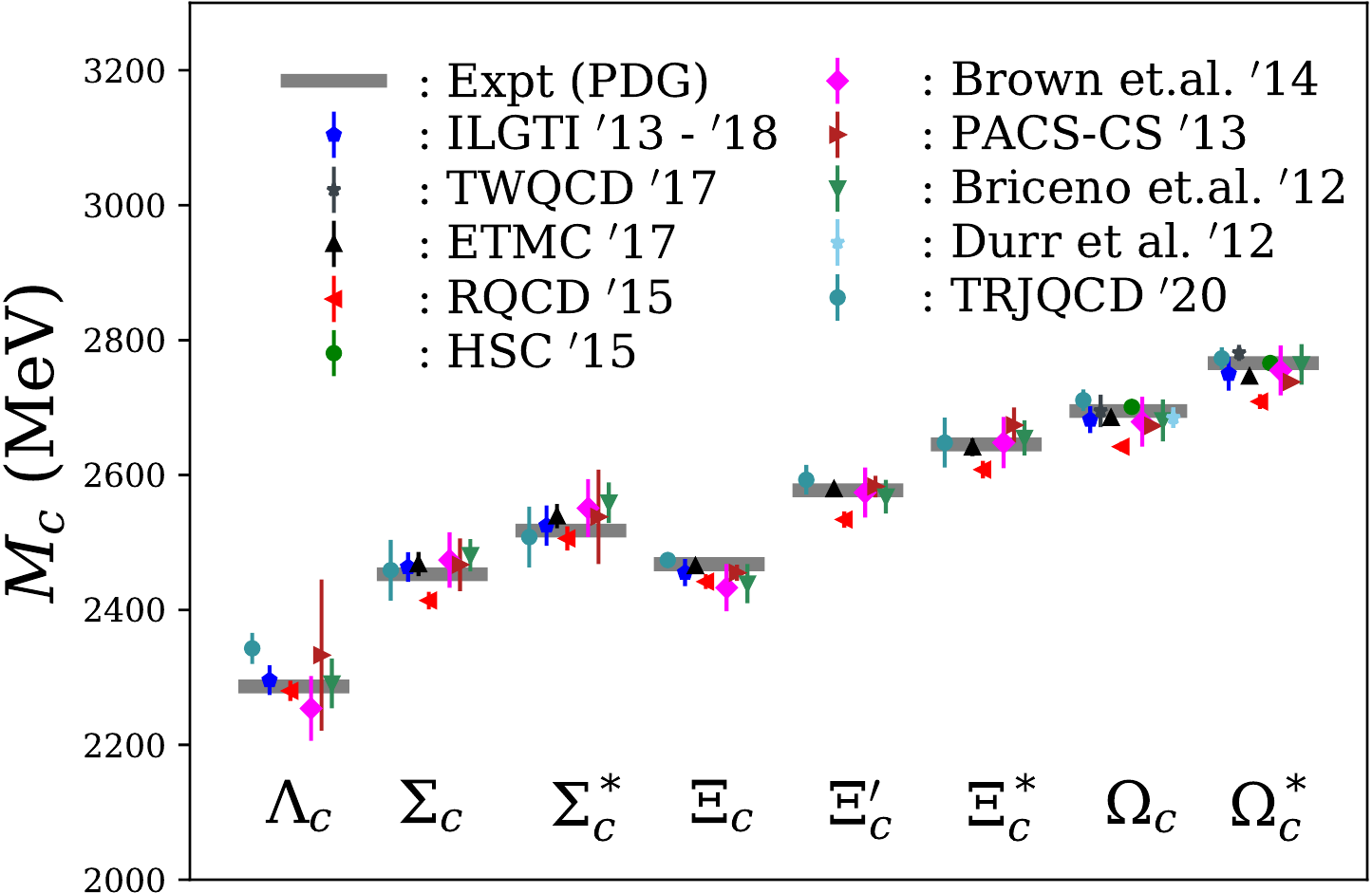} \qquad\qquad
\includegraphics[height=3.8cm,width=6.0cm]{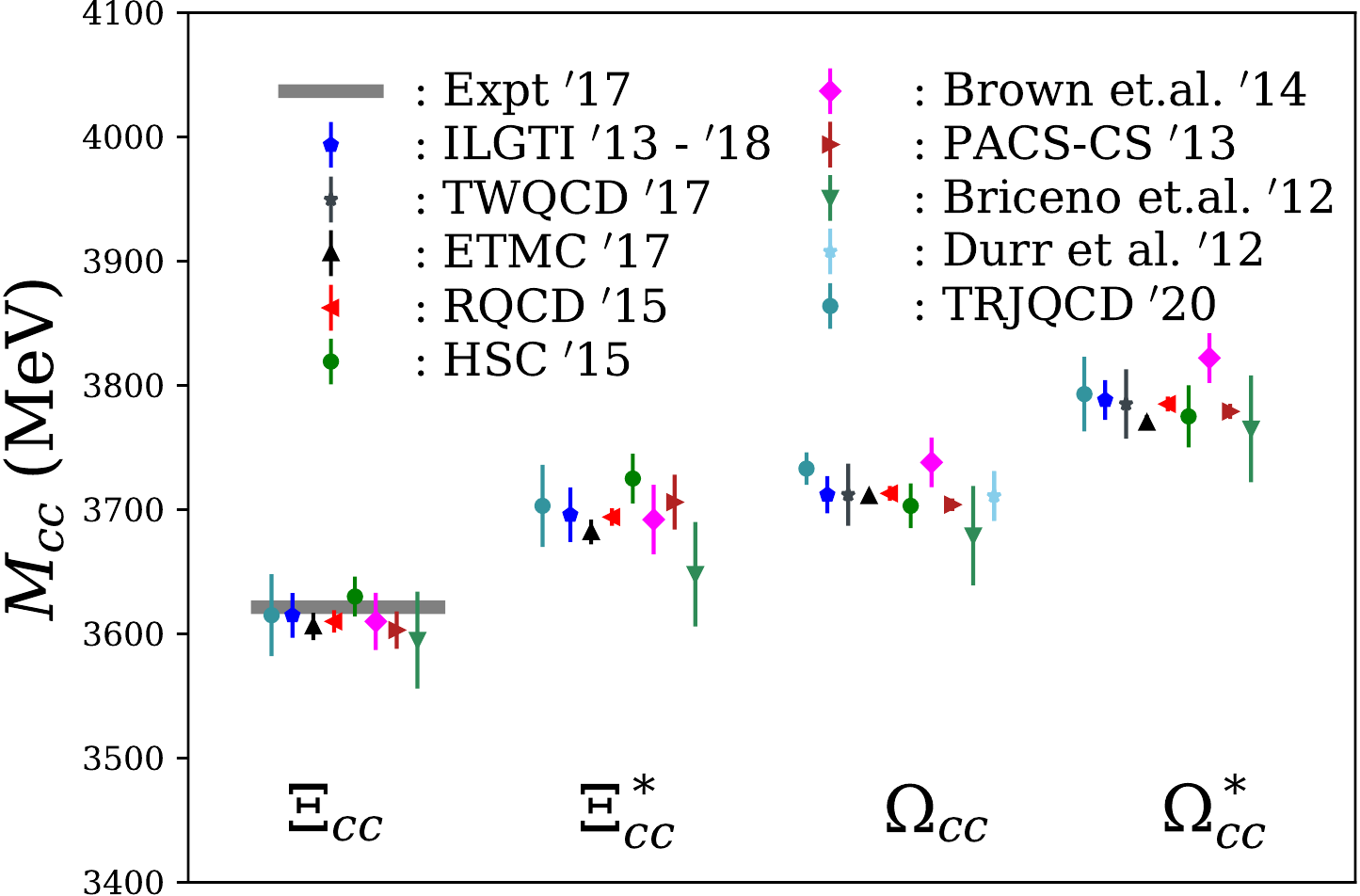}
\caption{Left : A summary of various lattice results for single 
charmed baryon masses. The horizontal lines represent the experimental 
masses. Right : Lattice predictions for the masses of the doubly charm 
baryons. The experimental mass for $\Xi_{cc}(1/2^+)$ as determined by LHCb
\cite{Aaij:2018gfl} is shown as a horizontal band.}\label{lcbaryons}
\end{figure}

Precise determination of the energy epectra of hadrons which are stable under strong 
interactions are benchmarks for the lattice investigations. There exist several 
calculations that have predicted/postdicted masses of heavy hadrons with full control 
over the statistical as well as systematic uncertainties (For mesons, {\it c.f.} 
Ref. \cite{DeTar:2018uko}). In Fig. \ref{lcbaryons}, we present the recent 
lattice results for ground-state masses for singly and doubly charmed baryons
(See Ref. \cite{Padmanath:2018zqw} for references).

{\bf Singly charmed baryons }: On the left side of Fig. \ref{lcbaryons}, we 
present a summary of recent lattice results for the masses of singly 
charmed baryons. The gray horizontal lines are experimental masses. 
Different lattice investigations utilize different setups and procedures, 
and hence have different systematics. In spite of this, an excellent overall 
agreement between all the lattice estimates and the respective experimental 
values is evident for all the hadrons. Note that in the physical world, 
$\Sigma_c$, $\Sigma^*_c$, and $\Xi^*_c$ can decay via strong interactions. 
Even so, the lattice estimates without any rigorous amplitude analysis agree 
remarkably well with the experiment values for these channels.

{\bf Doubly charmed baryons }: On the right side of Fig.~\ref{lcbaryons}, 
lattice predictions for the ground-state masses of doubly charmed 
baryons are presented. The only discovered doubly charmed baryon 
$\Xi_{cc}(1/2^+)$ \cite{Aaij:2018gfl} is presented by the horizontal 
line, which is consistent with all the lattice estimates. We 
emphasize that most of these lattice determinations were reported 
before the LHCb discovery and are all predictions. This demonstrates 
the potential of lattice QCD in making predictions in hadron 
spectroscopy calculations. The agreement between different lattice 
measurements indicates cut-off errors in these heavy hadron observables 
are small.

Another interesting point that we would like to mention is the 
comparison of lattice results with the SELEX candidate for a doubly 
charm baryon (3519(1) MeV) \cite{Mattson:2002vu}. All lattice 
estimates for $\Xi_{cc}(1/2^+)$ mass consistently lie $\sim 100$ MeV 
above the SELEX measurement. Furthermore, a lattice QCD+QED calculation 
with $N_f=1+1+1+1$ flavors performed by the BMW collaboration 
\cite{Borsanyi:2014jba} indicates the mass difference between 
the isospin partners of $\Xi_{cc}(1/2^+)$ to be $2.16(11)(17)$ MeV. 
Hence lattice investigations exclude any possibility of the SELEX 
measurement to be related to a doubly charmed baryon.

\begin{figure}
\includegraphics[height=3.8cm,width=6.0cm]{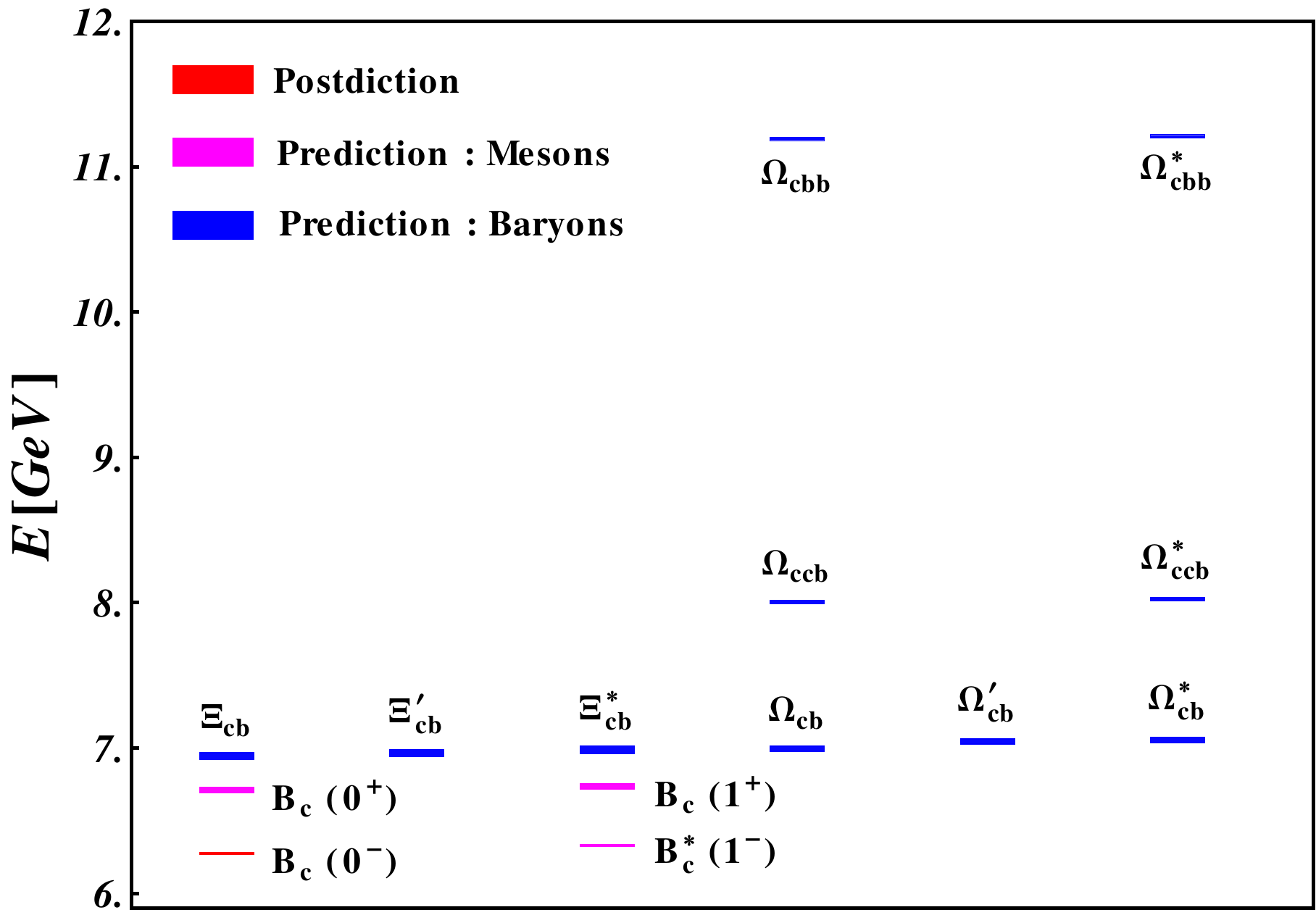} \qquad\qquad
\includegraphics[height=3.8cm,width=6.0cm]{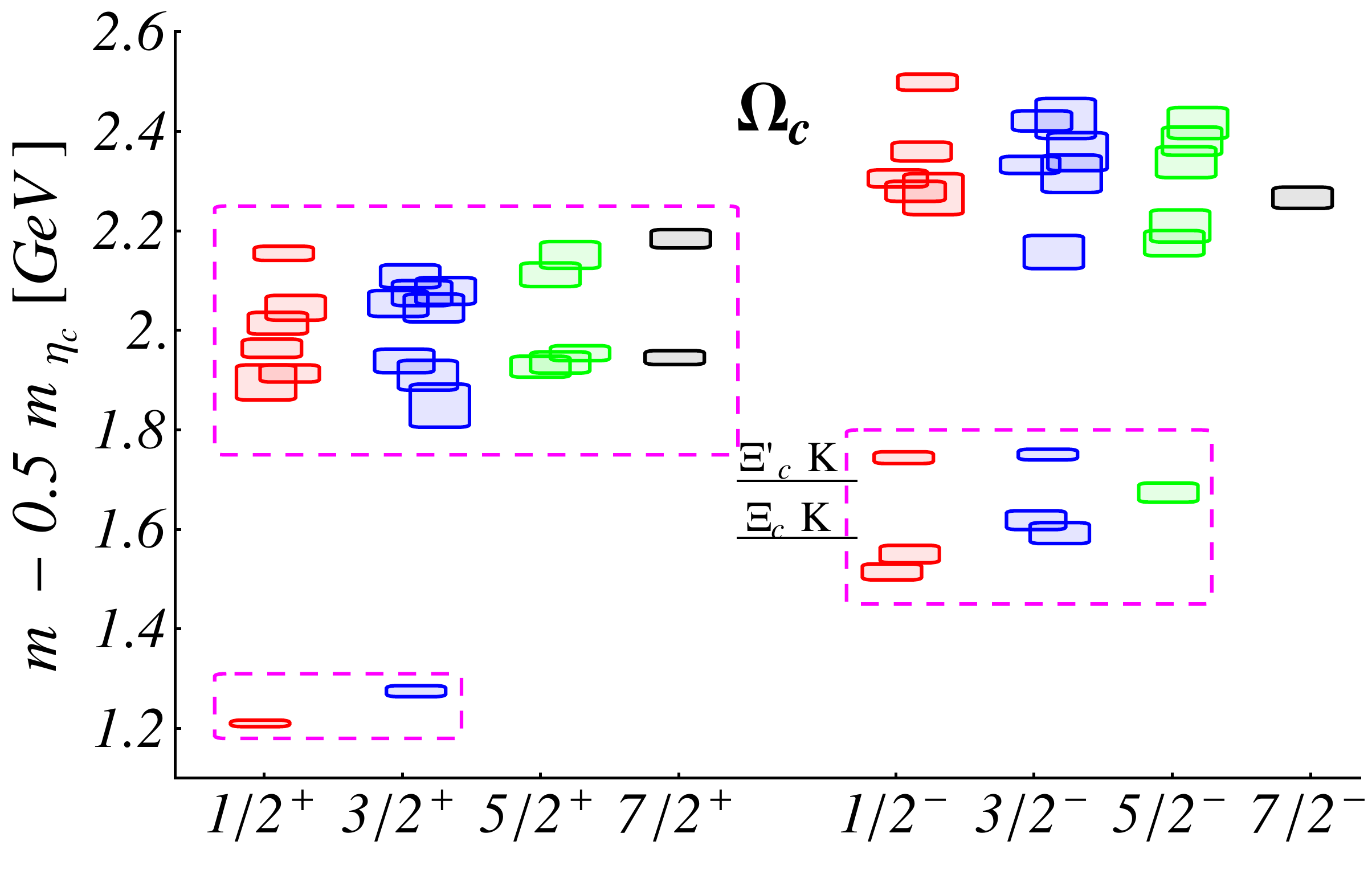}
\caption{Lattice estimates for charmed-bottom hadron masses as determined in Ref. \cite{Mathur:2018epb}. 
Excitation spectra of $\Omega_c$ baryon (right) as determined in Ref. \cite{Padmanath:2017lng}.}\label{bebaryons}
\end{figure}

{\bf Charmed bottom baryons }: On the left side of Fig. \ref{bebaryons}, we present 
lattice estimates for the ground-state masses of hadrons with at least one charm 
and one bottom quark studied on $N_f=2+1+1$ HISQ fermion MILC ensembles 
\cite{Mathur:2018epb}. In this investigation, controlled chiral and continuum 
extrapolations are performed to obtain reliable predictions for many 
yet-to-be-discovered charmed-bottom hadrons. The mass estimate for $B_c(1S)$ meson 
is consistent with the experimental value. The results for charmed bottom meson masses 
also agree with other existing lattice determinations \cite{Gregory:2010gm,Dowdall:2012ab}. 
The baryon mass estimates are also consistent with the only existing previous 
dynamical calculation \cite{Brown:2014ena} performed on ensembles with only two 
different lattice spacings.

{\bf Excited $\Omega_c$ baryons }: Not only the ground-state energy determination 
but hadron spectroscopy using lattice techniques has also made remarkable progress 
in determining the excitation spectrum. To this end, a basis of carefully 
constructed interpolating operators are used to compute the correlation matrices 
(Eq \ref{eq:2-2}) and variationally analyzed to extract the excitation energy spectrum. 
On the right side of Fig. \ref{bebaryons}, we present the excitation spectrum of 
$\Omega_c$ baryons in various $J^P$ quantum numbers \cite{Padmanath:2017lng}. 
Similar excitation spectra have also been determined using lattice QCD for other 
charmed baryons \cite{Padmanath:2013zfa,Padmanath:2015jea,Padmanath:2015bra} and 
triply bottom baryons \cite{Meinel:2012qz}. Given lattice results, the masses of 
recently discovered five excited $\Omega_c$ baryons by the LHCb Collaboration 
\cite{Aaij:2017nav} can immediately be compared.

\begin{figure}
\includegraphics[height=3.8cm,width=6.0cm]{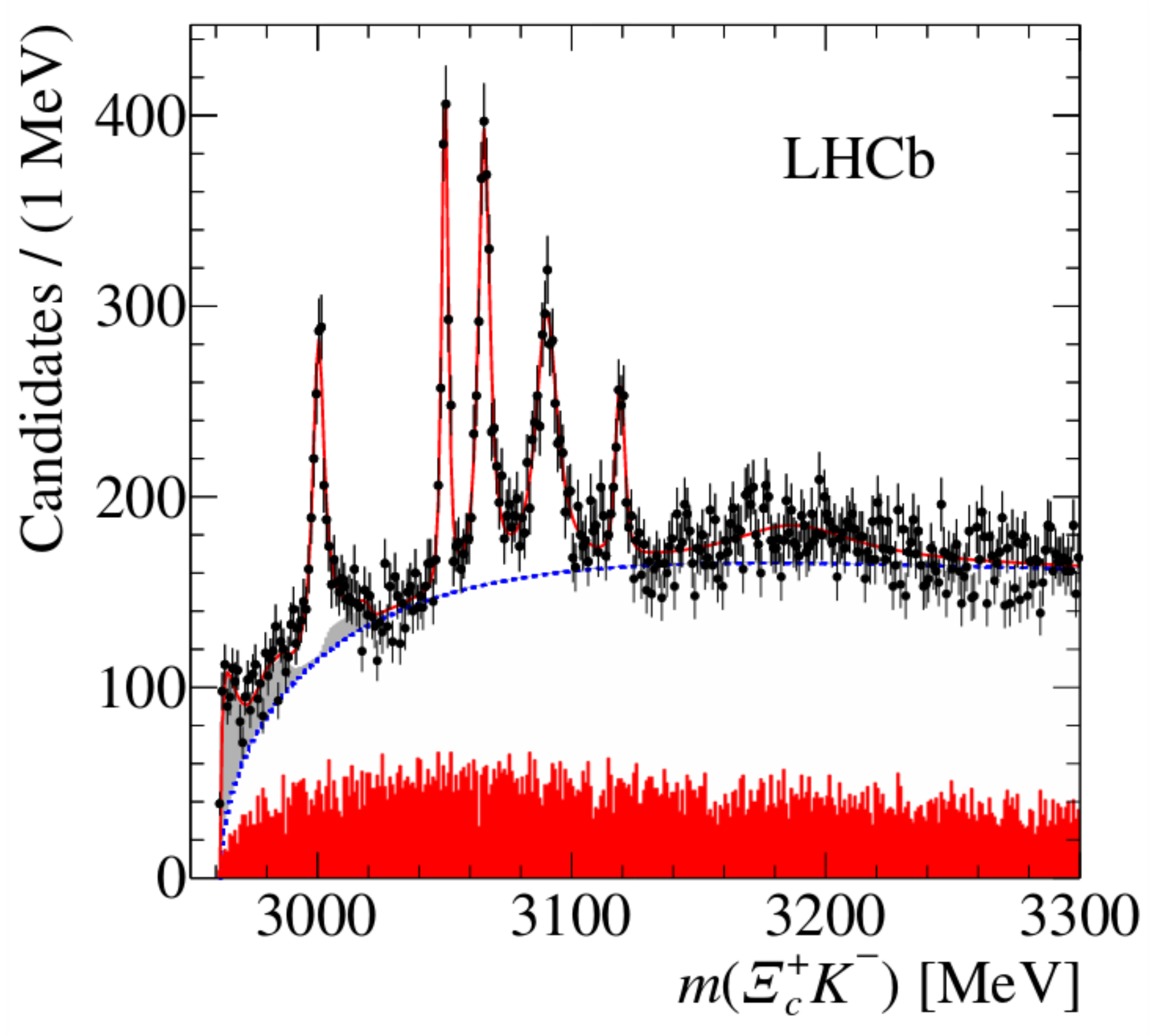} \hspace{2cm}
\includegraphics[height=3.8cm,width=5.2cm]{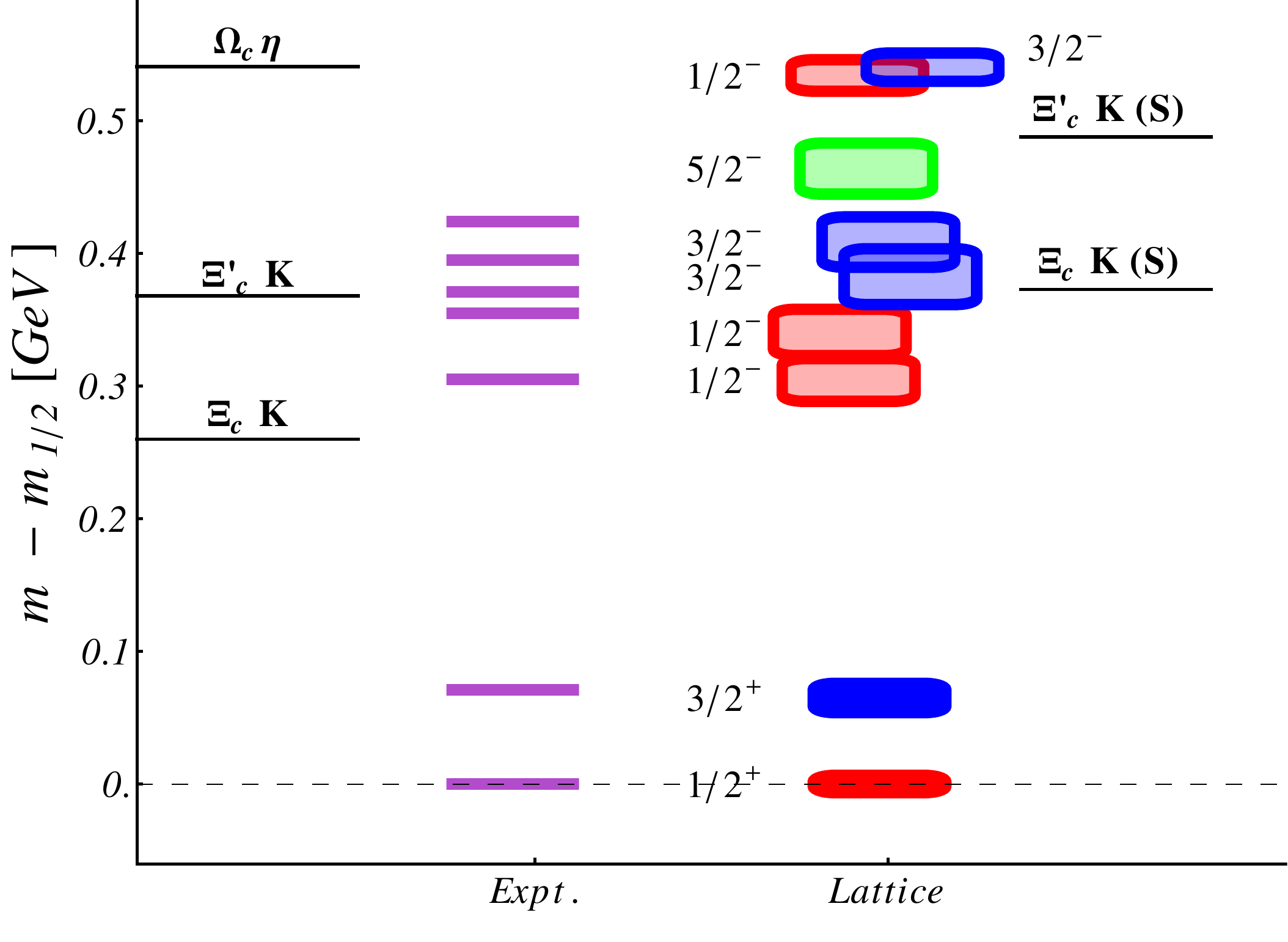}
\caption{Left : (Figure adapted from Ref. \cite{Aaij:2017nav}) The event distribution in the 
$\Xi_c^+K^-$ final states with peaks inferred as excited $\Omega_c$ baryons by the LHCb Collaboration. 
Right : The finite-volume energy spectrum is compared with the experimental masses of $\Omega_c$ 
baryons. Relevant low lying scattering thresholds are shown on the left and the finite-volume 
non-interacting energy levels are shown on the right as black horizontal lines. 
}\label{Ocbaryons}
\end{figure}

On the left side of Fig. \ref{Ocbaryons}, we show the discovery plot of excited $\Omega_c$ baryons 
by the LHCb Collaboration \cite{Aaij:2017nav} displaying the related five narrow peaks. 
The only existing lattice estimates for the excited $\Omega_c$ baryons \cite{Padmanath:2017lng} 
are compared with the experimental values on the right side of Fig.  \ref{Ocbaryons}. Working 
with a pion mass of $\sim$391 MeV and relatively coarse lattice spacing, this calculation aimed 
at an exploratory determination of the excitation spectrum and hence does not address various 
systematics in depth. The $1S$ hyperfine splitting in $\Omega_c$ baryons, which is known to be 
very sensitive to cut-off effects, can be seen to be consistent between the lattice 
estimate and the experimental value. Lattice QCD predicts five states in the energy range 
(0.3, 0.5) GeV, where the newly discovered $\Omega_c$ baryons lie. Qualitatively these five 
states correspond to the $p$-wave excitations of the $\Omega_c$ baryon. The authors of 
Ref. \cite{Padmanath:2017lng} also performed another lattice study with a different setup to check 
the extent of discretization uncertainties due to the coarse lattice spacing in the original
calculation. Although the second study did not use a large interpolator basis to extract all 
the low lying excitations, it utilized three lattice QCD ensembles with different lattice 
spacings to make a controlled continuum extrapolation of the ground-states they extract. The 
second lattice study confirmed the pattern of states qualitatively. These lattice results
were reported in several conferences such as Lattice 2014 \cite{Padmanath:2014bxa} as well as 
in Charm 2013, 2015 \cite{Padmanath:2013bla,Padmanath:2015bra} and hence preceded the LHCb 
discovery. Recent LHCb report conflicts these quantum number assignments to these five 
states \cite{LHCb:2021ptx}. This indicates the necessity of systematic lattice investigations 
involving the amplitude analysis on finite-volume spectrum using finer lattices to resolve 
the discrepancy quantitatively.

An obvious next step towards this is to include multi-hadron interpolators that are related 
to the nearby non-interacting baryon-meson levels and rigorous finite-volume analysis. Lattice 
calculations using baryon-meson interpolators are limited to a few. Most of them are in the 
light baryon sector (For more info see Ref. \cite{Padmanath:2018zqw})). More recently, two 
studies investigated interactions in charmonium-nucleon systems \cite{Alberti:2016dru,
Skerbis:2018lew}. In Ref. \cite{Alberti:2016dru}, the authors studied the static $Q\bar Q$ 
potential in the presence of various light hadrons, including baryons. In all the cases, they 
found that bound state configurations are energetically favored. In Ref. \cite{Skerbis:2018lew}, 
the authors reported the finite-volume spectra for nucleon-$J/\psi$ and nucleon-$\eta_c$ systems, 
and their results did not indicate any possibility for bound states or resonances in these 
channels. More lattice calculations of the excited baryon spectrum will be highly appreciated by 
the scientific community, anticipating the discovery of many more baryons in experiments 
like LHCb and Belle.

\section{Heavy exotics}\label{beyond}

\begin{figure}
\centering
\includegraphics[height=4.0cm,width=4.8cm]{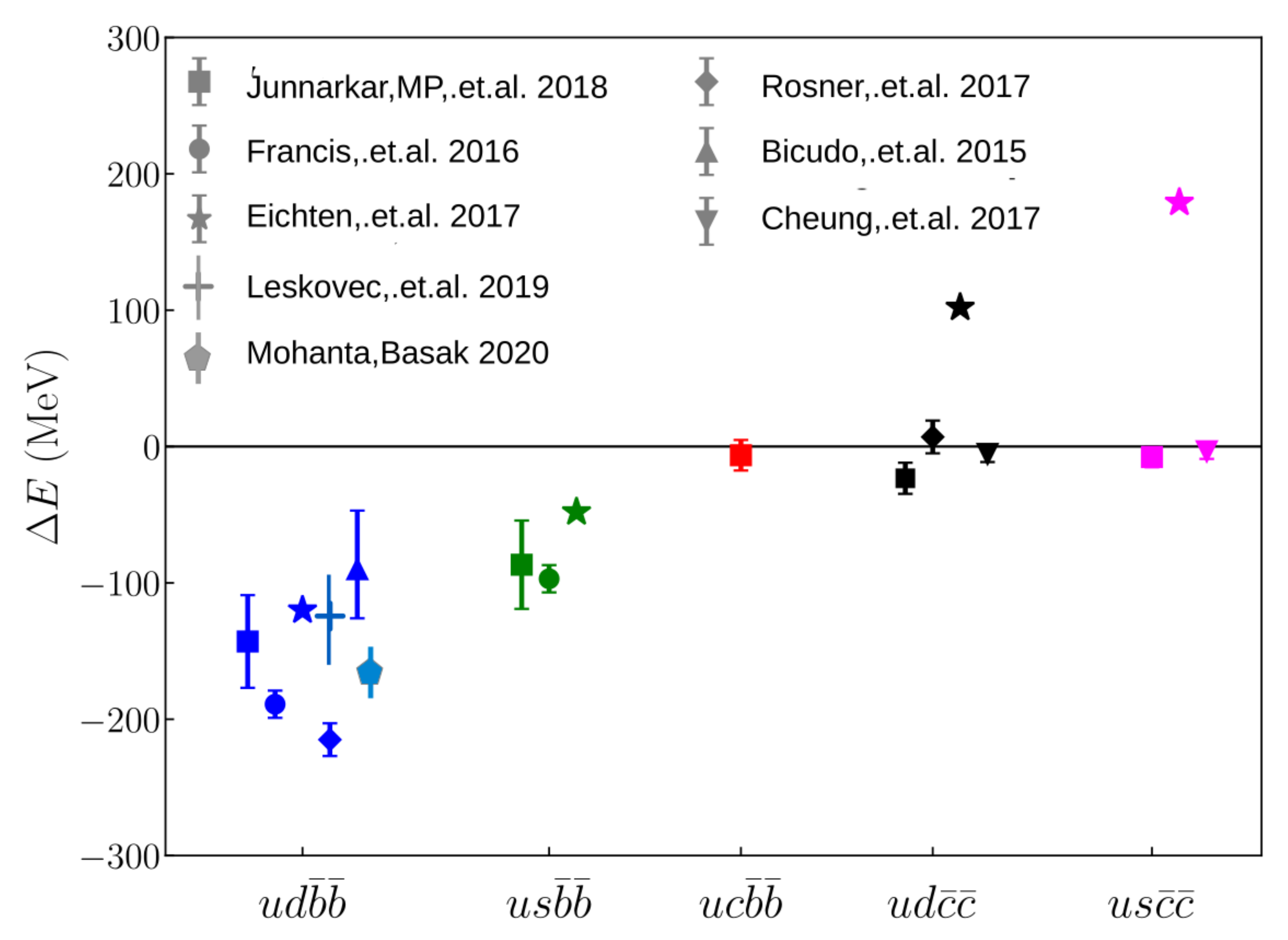} 
\includegraphics[height=4.0cm,width=4.8cm]{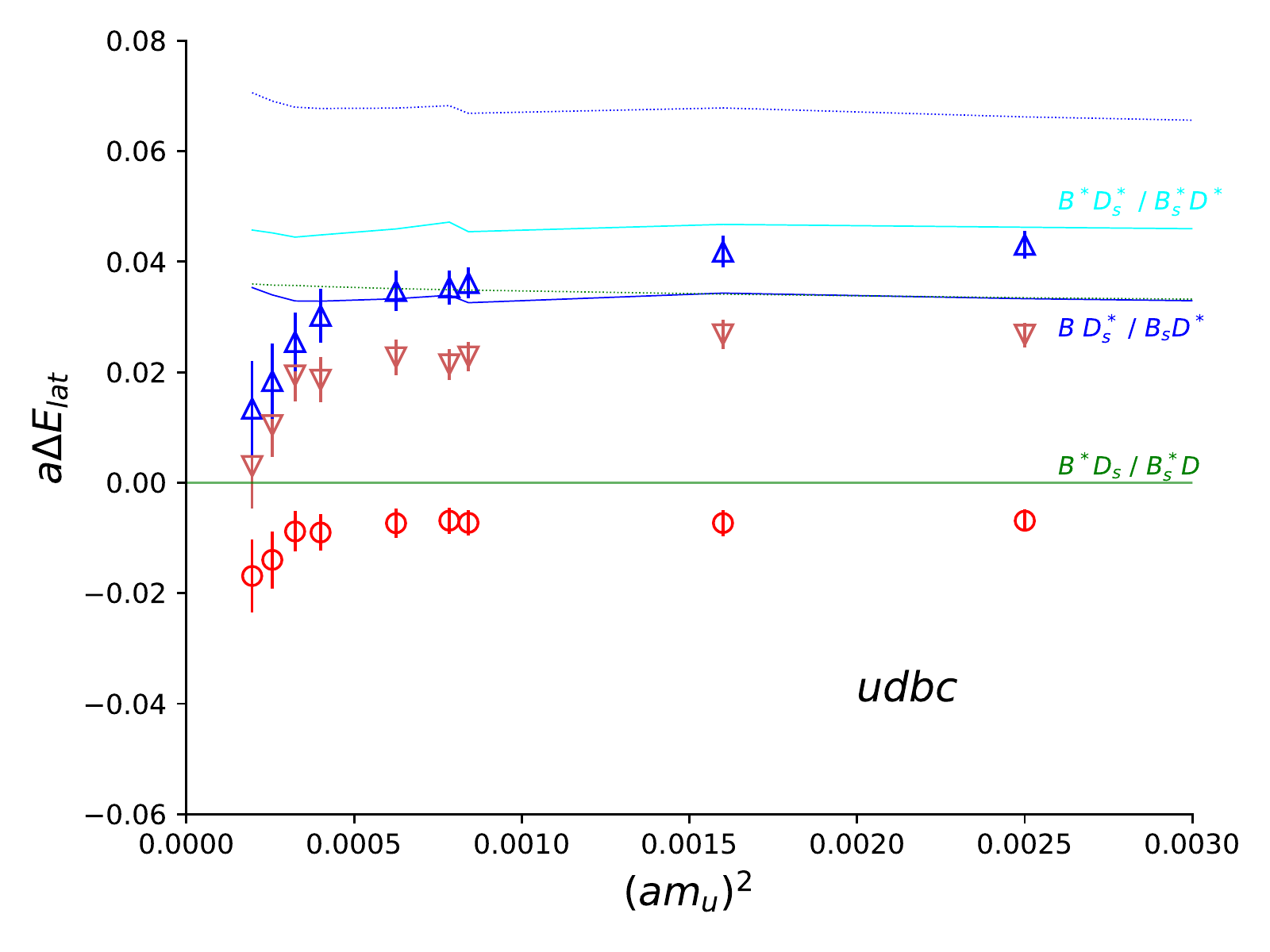}
\includegraphics[height=4.0cm,width=5.2cm]{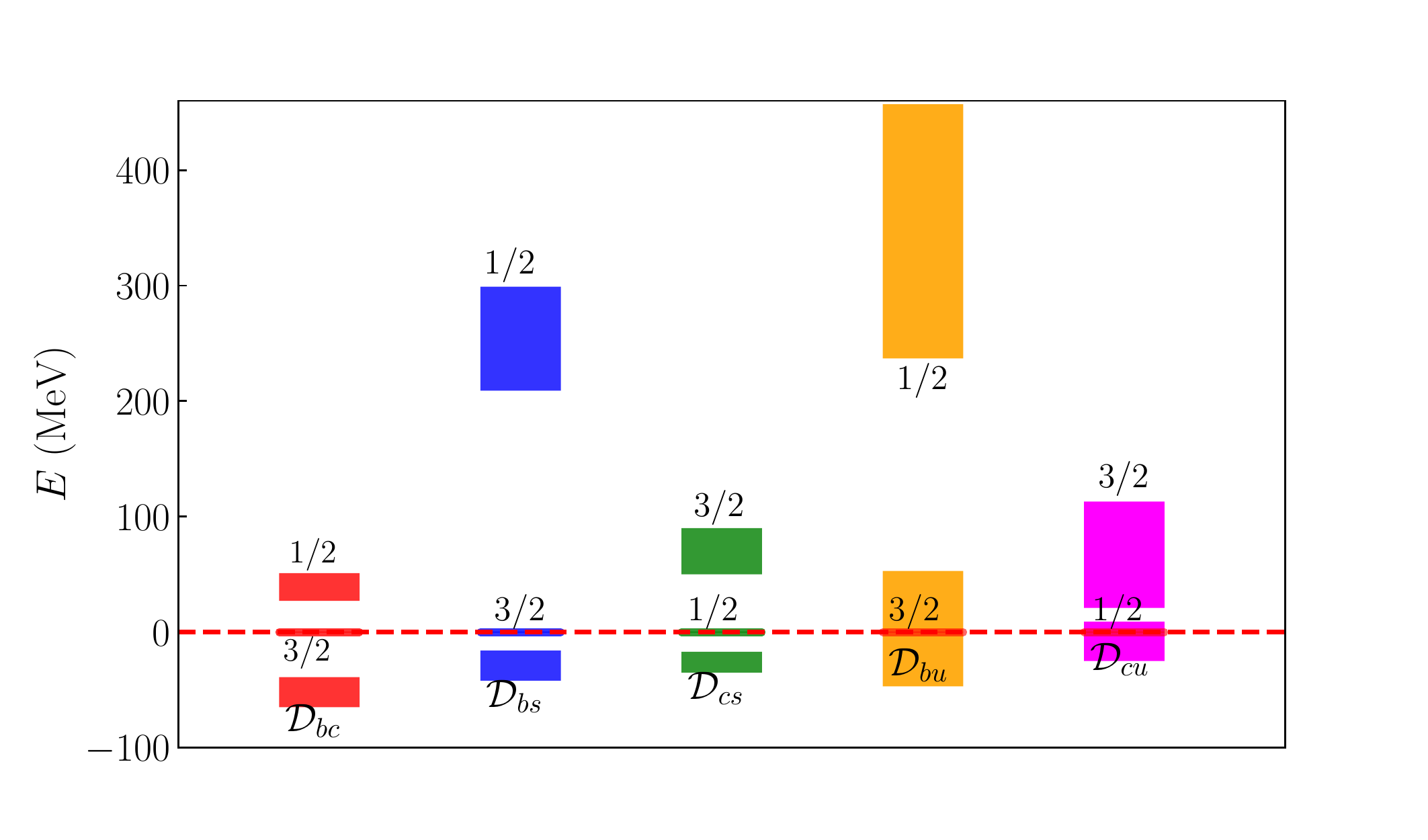}
\caption{Left: Summary of all the lattice results and a selected set of non-lattice estimates for 
the ground-state energies in the doubly heavy tetraquark channels. Center: The light quark mass 
dependence of the finite-volume spectrum for the $\bar b\bar cdu$ four quark system by the ILGTI collaboration. 
Right: Low lying finite-volume spectra of Deuteron-like heavy dibaryons as determined
in Ref. \cite{Junnarkar:2019equ}}\label{bbtq}
\end{figure}

{\bf Doubly bottom and doubly charm tetraquarks}:
Early model calculations indicated that doubly heavy four quark systems in the heavy 
quark limit to be a promising platform to find stable tetraquark states \cite{Carlson:1987hh}. 
This has received significant attention over the recent years both in phenomenological 
studies \cite{Karliner:2017qjm,Eichten:2017ffp} and on the lattice \cite{Bicudo:2016ooe,
Bicudo:2017szl,Francis:2016hui,Junnarkar:2018twb,Francis:2018jyb,Leskovec:2019ioa,
Hudspith:2020tdf,Mohanta:2020eed}. A summary of various lattice and some non-lattice 
calculations can be found on the left side of Fig. \ref{bbtq}. Lattice investigations remain 
exploratory in this regard, and except one, most lattice results are based on the 
ground-state mass estimates. Even though this is the case, qualitative inferences can 
be made for possible bound state scenarios. Only in Ref. \cite{Leskovec:2019ioa}, the 
authors perform an amplitude analysis to determine the actual binding energy, 
whereas in Refs. \cite{Bicudo:2016ooe,Bicudo:2017szl} the authors determine static quark 
potentials in the presence of two light quarks on the lattice and search for bound state 
poles. In the doubly bottom axial-vector channel, all lattice investigations predict a 
deeply bound state with binding energy $\mathcal{O}(100 MeV)$ for both $I=1$ 
($\bar b\bar bdu$) and $I=1/2$ ($\bar b\bar bsu$). For $\bar b\bar bcu$, the lowest 
finite-volume level was found to be consistent with the threshold \cite{Junnarkar:2018twb}. 
Similar investigations in the doubly charm sector have suggested the lowest finite-volume 
level immediately below or consistent with the elastic threshold. The $cc\bar d\bar u$ 
channel is related to the $T_{cc}$ reported by LHCb recently. A chiral and 
continuum extrapolated estimate for $\bar c\bar cdu$ ground-state in the finite-volume 
by ILGTI collaboration suggests possible binding energy of up to 23 MeV 
\cite{Junnarkar:2018twb}. However, a rigorous amplitude analysis of the finite-volume 
spectrum is highly desired to assess such near-threshold poles quantitatively.  

{\bf Charmed-bottom tetraquarks}:
It is also interesting to investigate possible scenarios if one of the heavy quarks 
is replaced with a lighter heavy quark. Investigations in these systems have already 
been reported in Refs. \cite{Francis:2018jyb,Hudspith:2020tdf}, which in conclusion 
claimed to observe no finite-volume levels below the elastic thresholds in any channels 
other than in the doubly bottom four quark systems. These investigations were performed 
on PACS-CS lattices with single lattice spacing ($\sim 0.09$ fm). A recent lattice 
investigation by ILGTI collaboration utilizing up to three lattice spacings ($\sim 0.12$ 
fm, $\sim 0.09$ fm, and $\sim 0.06$ fm) suggests that, on the fine lattice spacing, 
they find finite-volume levels that are quantitatively below the elastic threshold. 
They also observe that the interpolator basis commonly used for all these calculations 
is based on constraints in the heavy quark limit, whereas away from this limit, other 
allowed operators could be essential to determine the complete excitation spectrum 
reliably. In the center of Fig. \ref{bbtq}, we present the light quark mass dependence 
of the finite-volume spectrum in the $\bar b\bar cdu$ axial-vector channel [in the 
range of 250 MeV - 800 MeV pion mass] on the finest lattice. At all quark mass values, 
there is at least one level below the elastic threshold, unambiguously pointing to 
interesting physics near the threshold. The zero in the y-axis refers to the elastic 
threshold. 

{\bf Heavy dibaryons }:
Several lattice groups have investigated dibaryon spectra. However, all were focussed 
on the light/strange dibaryons {\it c.f.} Refs. \cite{NPLQCD:2010ovm,Inoue:2010es,
Green:2021qol}. In a recent investigation \cite{Junnarkar:2019equ}, the authors 
have extracted the finite-volume spectrum of Deuteron-like heavy dibaryon systems 
looking for evidence of possible bound states in such system. Given that the Deuteron 
is a bound state in the physical world, bound states are naturally expected in its 
heavy analogues. The spectrum for a variety of dibaryon channels with flavor 
patterns $\Sigma_c\Xi_{cc}$,  $\Omega_c\Omega_{cc}$, $\Sigma_b\Xi_{bb}$, 
$\Omega_b\Omega_{bb}$, and $\Omega_{ccb}\Omega_{cbb}$ in the axial-vector channel 
were reported in the article. They find strong indications for bound states in 
the dibaryon channels with $\Omega_h$ baryons, while robust inferences could not be 
made due to large systematic uncertainties in the other two channels. A summary plot 
of their findings is shown on the right side of Fig \ref{bbtq}.

Except for a few calculations, all other lattice results presented above ignore 
the effects of any nearby strong decay thresholds. This is justified for most of the baryons 
discussed in the previous section. However, for the four-quark systems discussed in 
this section and the $\Sigma_c$, $\Sigma_c^*$ and $\Xi_c^*$ baryons that can decay via 
strong interactions, such a procedure is questionable. In these cases, a rigorous 
determination of the excitation spectra followed by a proper finite-volume analysis, 
such as discussed in the next section, is desired.

\section{Excited and exotic charmonium}\label{mesres}

A large fraction of experimentally known heavy quark exotics is found in the charmonium 
spectrum. These are referred to as XYZs and have been discovered with properties 
contradicting the theoretical expectations from simple minded potential models.  
The first one in this family is $X(3872)$ discovered by Belle in 2003 \cite{Choi:2003ue}. 
Today, there are several such candidates with unexplained nature in the charm and the bottom 
sectors. A summary of efforts to find a theoretical description of these states can be 
found in Refs. \cite{Esposito:2016noz,Olsen:2015zcy}.

Many precision lattice calculations have been made for the ground-state charmonium masses, 
{\it c.f.} Ref. \cite{DeTar:2018uko} and the references therein. Several groups have also 
determined the excited charmonium spectrum in the finite-volume \cite{Bali:2011rd,Liu:2012ze,Mohler:2012na,
Padmanath:2018tuc}. However, rigorous investigations of the relevant strong interaction 
thresholds and scattering amplitudes are limited to a few. Two lattice calculations extracted 
a bound state pole associated with $X(3872)$ assuming $D\bar D$ elastic scattering 
\cite{Prelovsek:2013cra,Padmanath:2015era}. These works assumed the effects of couplings with 
other open scattering channels to be negligible. A similar investigation in the hidden charm 
$I=1$ sector indicated no signatures for the existence of any interaction in the low energy 
regime \cite{Prelovsek:2013xba}. Recently HadSpec Collaboration has performed a detailed calculation to extract 
the finite-volume spectra in the rest frame for $I=1$ hidden charm \cite{Cheung:2017tnt}. 
They also arrived at a similar inference from their finite-volume spectra as in Ref. 
\cite{Prelovsek:2013xba}. All of these lattice calculations were limited to the rest 
frame. In another letter by the HALQCD Collaboration, using their finite-volume formalism, 
they investigate the interactions between $\pi J/\psi$, $\rho\eta_c$ and $\bar DD^{\ast}$ 
and argues the charged $Z_c(3900)$ to be a threshold cusp \cite{Ikeda:2016zwx}.

In the vector charmonium spectrum, there are two bound states $J/\psi$, $\psi(2S)$ and 
the $\psi(3770)$ resonance below 4 GeV. Given the sparse spectrum and the elastic nature of the 
$\psi(3770)$ resonance, this channel can be studied as the proof of principle. In 
the scalar charmonium, the situation is much less clear with a single bound state 
$\chi_{c0}(1P)$ and three exotic candidate resonances around the energy region 3.9 GeV \cite{Chilikin:2017evr,
Aaij:2019evc}. Only one lattice investigation was performed, till 2019, for these channels 
assuming an elastic scattering of $D\bar D$ in the rest frame \cite{Lang:2015sba}. In what 
follows, we will discuss a recent calculation of these channels performed by the RQCD 
Collaboration in an extended lattice QCD setup. This new calculation utilizes two 
lattice QCD ensembles and works in three different inertial frames to extract the 
physics. The ensembles used have $m_{\pi}\sim280$ MeV and $m_K\sim 467$ MeV, and with 
spatial extents $L\sim2$ fm and $L\sim2.7$ fm. The calculation was performed at two 
different values of charm quark masses to investigate the influence of the open-charm threshold.

\begin{figure}
\centering
\includegraphics[height=4.6cm,width=7cm]{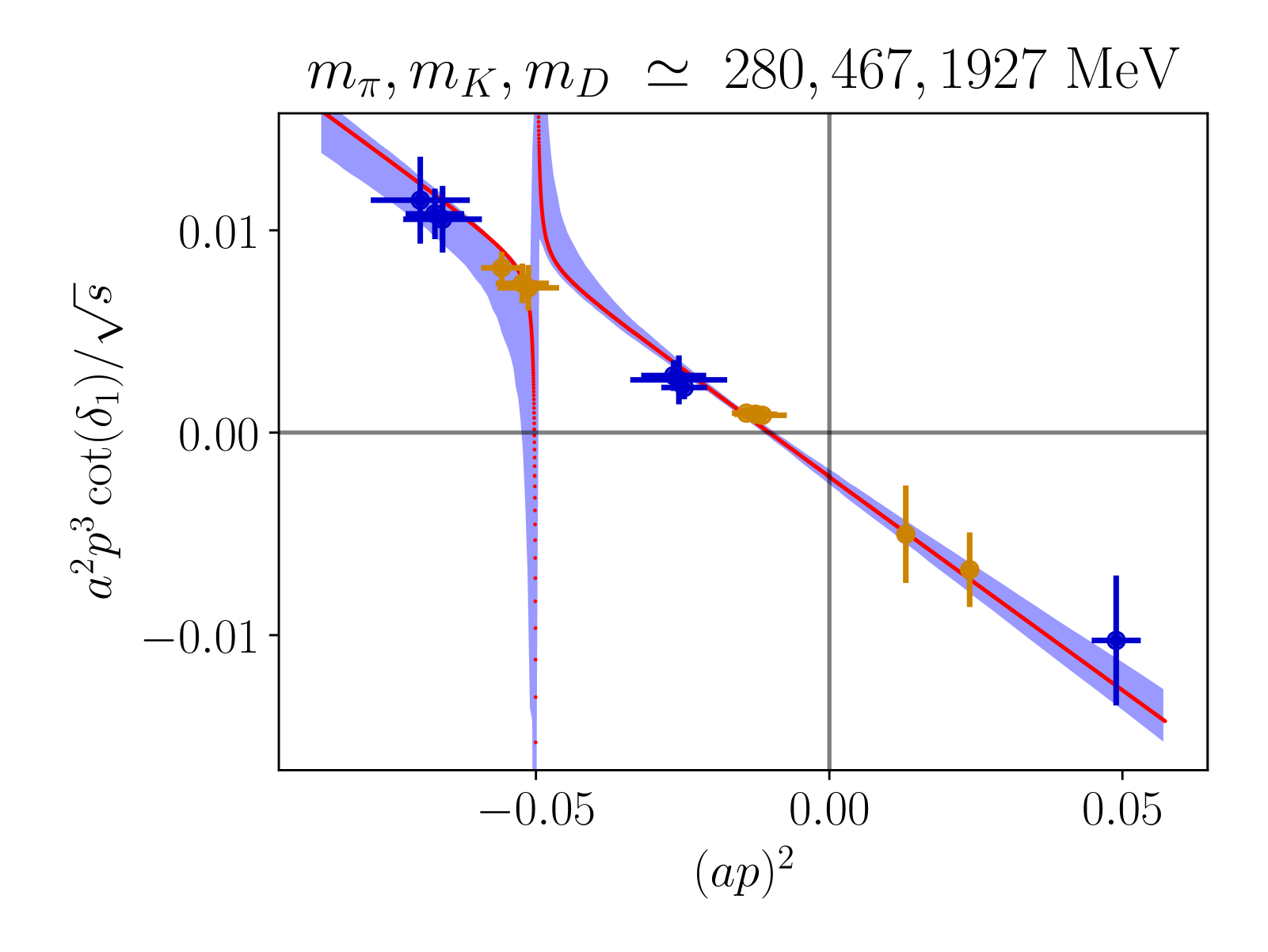}
\includegraphics[height=4.6cm,width=7cm]{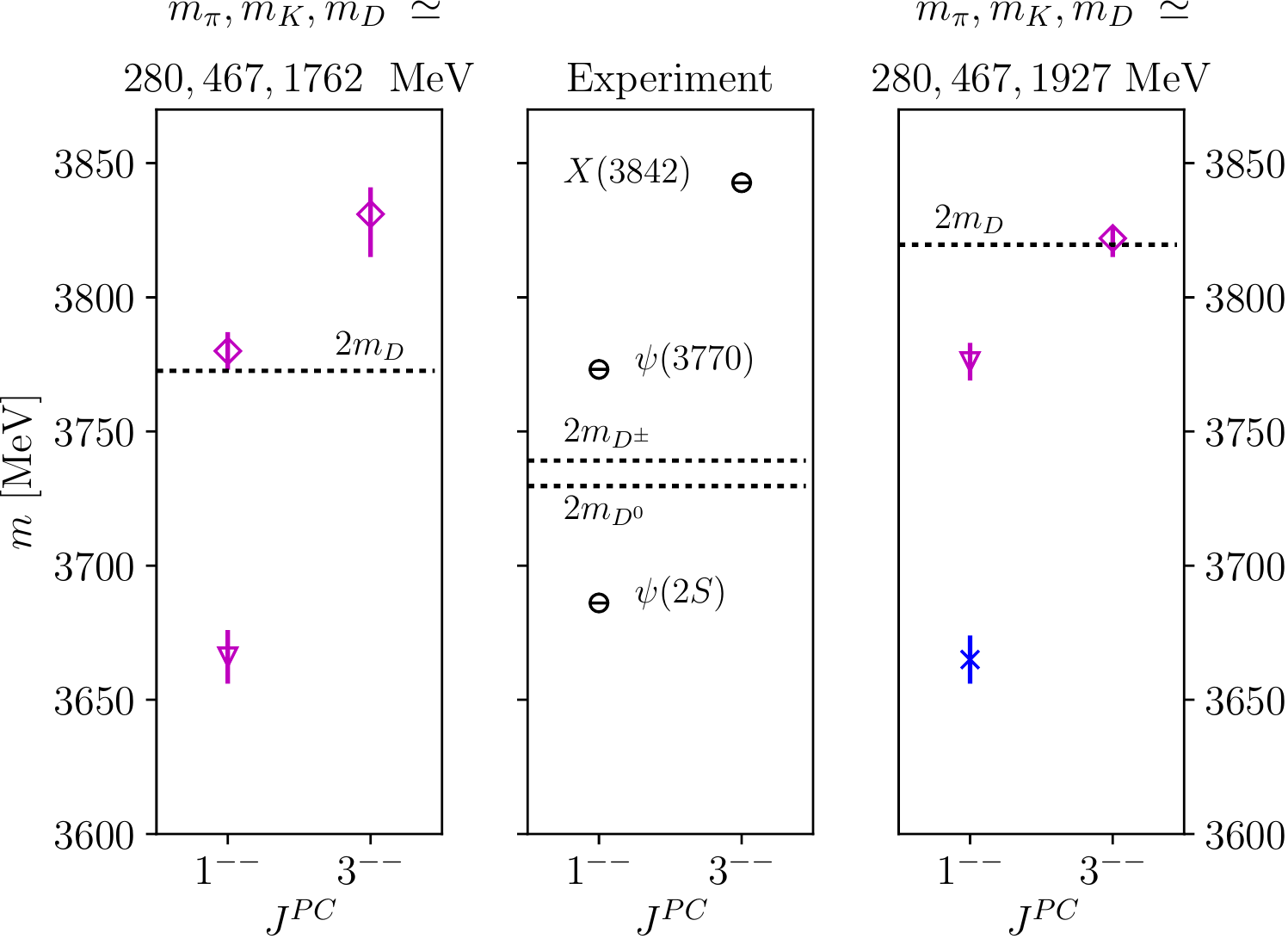}
\caption{ (Figures adapted from Ref. \cite{Piemonte:2019cbi}) Left: $D\bar D$ scattering amplitude in the vector 
channels as a function of the square of scattering momentum. Results are shown for the heavier than physical 
charm quark mass. Right: The infinite-volume spectrum extracted from the finite-volume spectra are compared with 
the respective experimental values.}
\label{rqcdvector}
\end{figure}

In Figure \ref{rqcdvector}, we present the results obtained for the vector charmonium 
spectrum in $D\bar D$ scattering in $l=1$ partial wave. Each finite-volume energy 
level can be associated with an infinite-volume phase shift for elastic scattering 
of spinless particles in a single partial wave. On the left, the $D\bar D$ scattering 
phase-shifts, ($a^2p^3cot(\delta_1)/\sqrt{s}$), are plotted as a function of the 
scattering momentum. The data points represent finite-volume spectra along the x-axis 
and the $D\bar D$ scattering phase-shifts along the y-axis. The extracted $D\bar D$ 
scattering phase-shift is then fit with double pole parametrization of the scattering 
matrix, represented by the solid red curve and the associated errors by the blue band. 
The pole positions and residues of the singularities in the parametrized scattering 
matrix in the complex s-plane are extracted by analytic continuation. On the right, we present 
the extracted infinite-volume spectra along with the respective experimental numbers. 
It is evident from the plots that the extracted energies of the excitations remain mostly 
unaffected with unphysical threshold positions and are consistent with the experimental values.
 
\begin{figure}
\centering
\includegraphics[height=4.0cm,width=4.5cm]{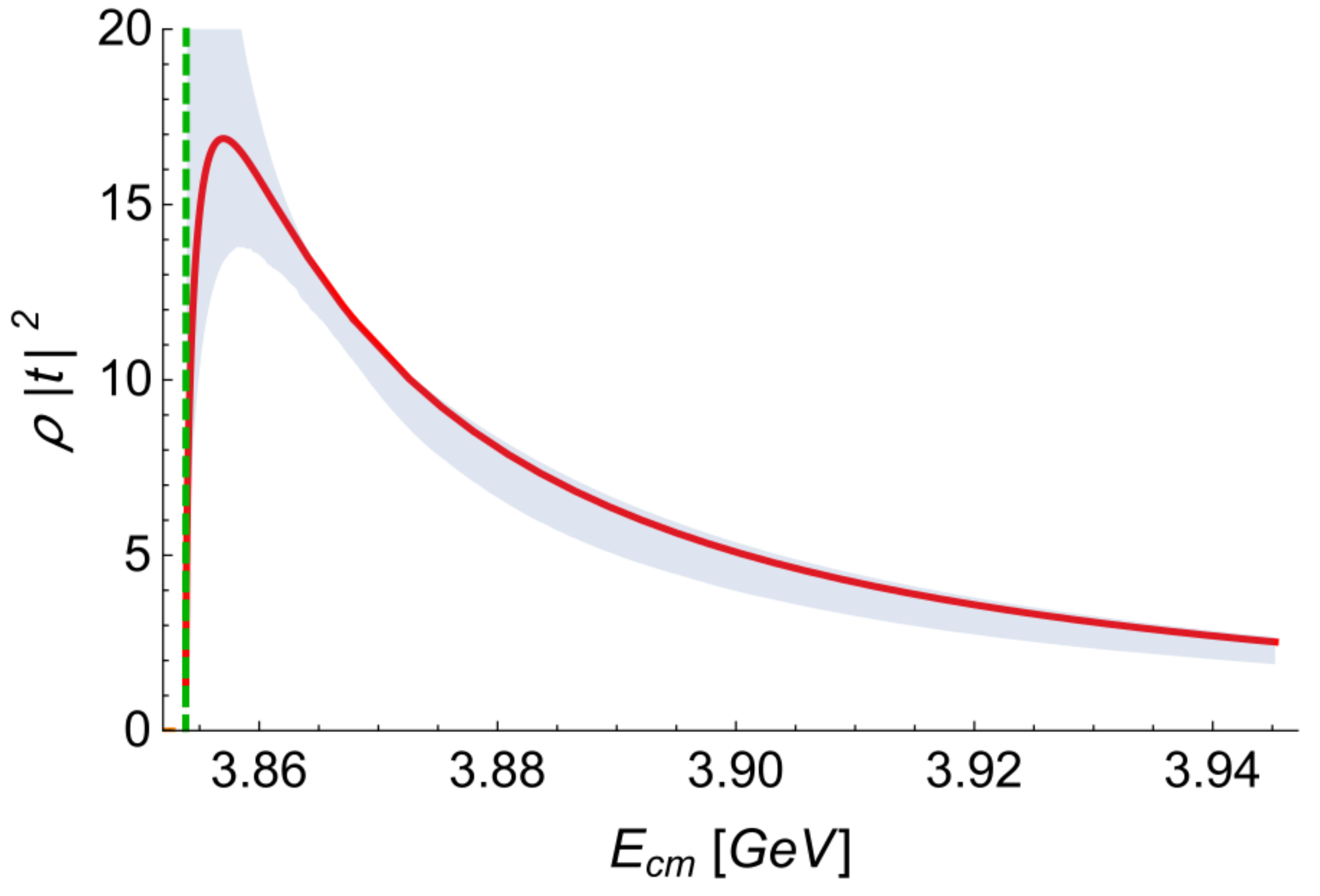}
\includegraphics[height=4.0cm,width=4.5cm]{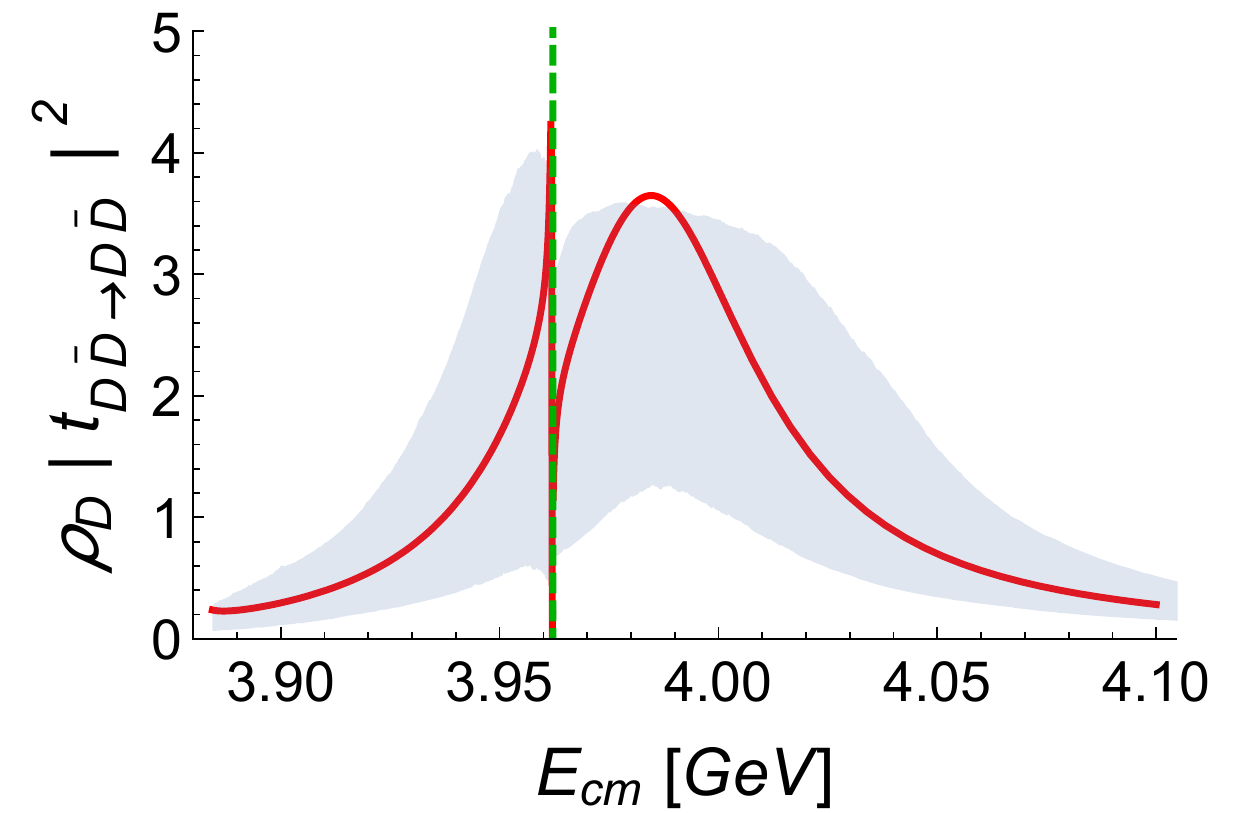}
\includegraphics[height=4.0cm,width=4.5cm]{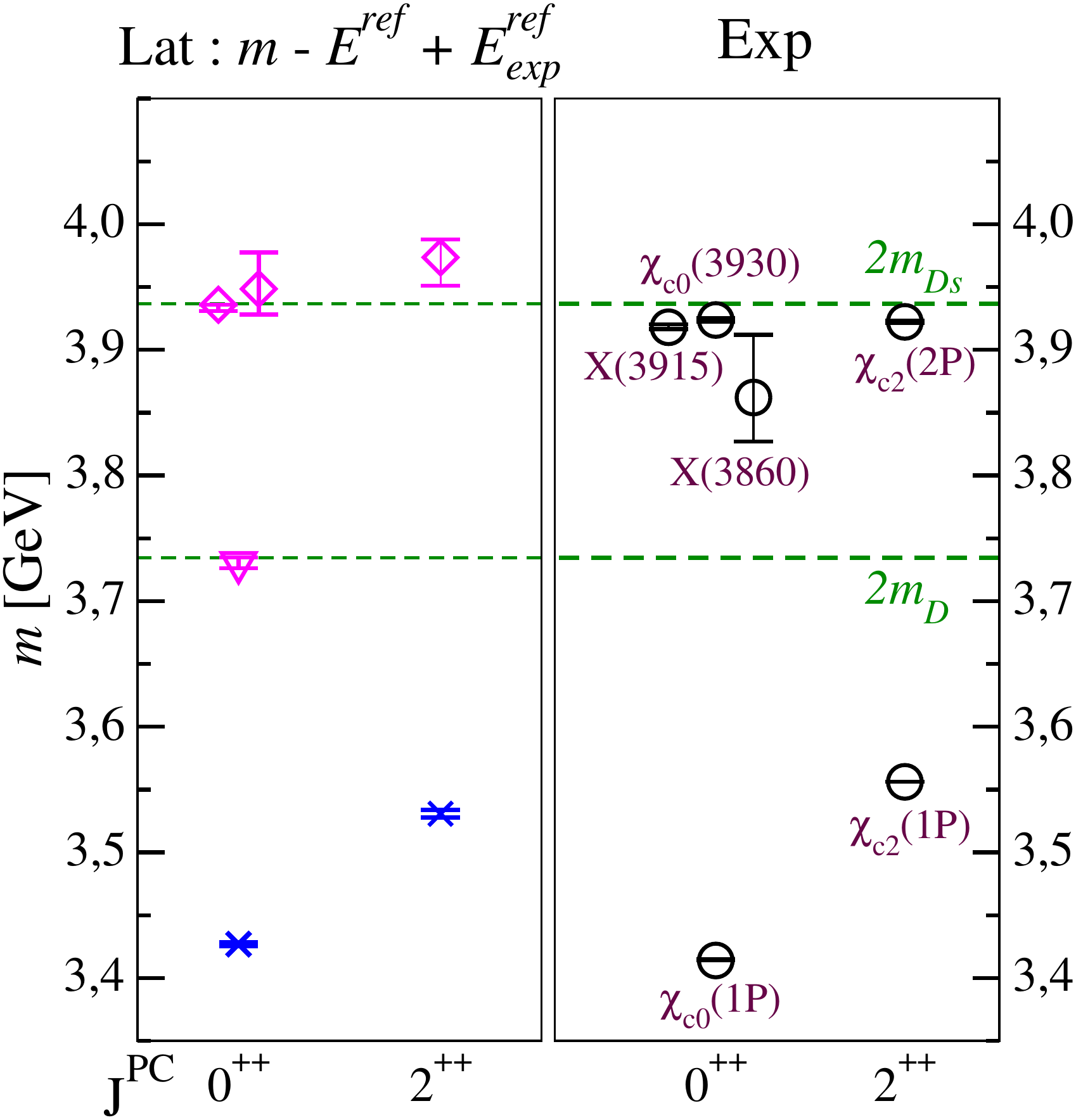}
\caption{ (Figures adapted from Ref. \cite{Prelovsek:2020eiw}) Left and center: $D\bar D$ scattering amplitude 
in the scalar channel as a function of the center of momentum energy, $E_{cm}$. On the left, the amplitude 
close to the $D\bar D$ threshold (vertical green dashed line) is shown. In the center, the amplitude in 
the energy region around $D_s\bar D_s$ threshold (vertical green dashed line) is plotted. Right: The extracted 
charmonium spectrum is compared with the experimental values.}
\label{rqcdscalar1}
\end{figure}

In Figure \ref{rqcdscalar1}, we present the results from the lattice investigation of 
scalar charmonium spectra in $D\bar D$ and $D_s\bar D_s$ multi-channel scattering. This 
investigation considers the energy region with a lower bound slightly below the $D\bar D$ 
threshold and an upper bound around 4.1 GeV. In the energy region around the open charm 
elastic threshold ($D\bar D$), a very shallow bound state pole is required to describe 
the observed finite-volume spectrum. The presence of such a pole is expected to cause 
a rapid rise in the $D\bar D$ scattering amplitude immediately above the threshold, as 
shown in the left plot in Figure \ref{rqcdscalar1}. Such a shallow bound state in 
$D\bar D$ channel was also proposed by phenomenological models such as in Refs. 
\cite{Gamermann:2006nm,Hidalgo-Duque:2013pva,Baru:2016iwj}.

In the higher energy region starting from slightly below the $D_s\bar D_s$ threshold, 
the authors argue the presence of two pole singularities in the extracted scattering 
amplitude across the complex s-plane, that can have signatures on the real axis. One pole 
exists above the $D_s\bar D_s$ threshold in the Riemann sheet III (-,-) with large 
coupling to the $D\bar D$ channel. Such a pole is expected to reflect in the $D\bar D$ 
scattering amplitude as a vanilla resonance peak (see the center plot in Figure 
\ref{rqcdscalar1}). Considering the extracted mass and its strong coupling to the 
$D\bar D$ scattering channel, the authors relate this state to the $\chi_{c0}(3860)$ 
discovered recently by Belle \cite{Belle:2017egg}.

Another pole is observed very close below the $D_s\bar D_s$ threshold in the Riemann 
sheet II with large coupling to $D_s\bar D_s$ channel. As a result of the interference 
with the pole above the $D_s\bar D_s$ threshold, this pole features in the $D\bar D$ 
amplitude as a prominent dip falling to zero below the $D_s\bar D_s$ threshold. This 
is also visible in the center plot in Figure \ref{rqcdscalar1}. This dip in the 
$D\bar D$ crosssection is very similar to the dip in the $\pi\pi$ crosssection slightly 
below the $K\bar K$ threshold related to the $f_0(980)$ resonance \cite{Briceno:2017qmb}. 
Except at the $D_s\bar D_s$ threshold, where a kink appears, the $D\bar D$ amplitude 
remains smooth and continuous. The presence of this pole also causes a rapid rise in 
the $D\bar D\rightarrow D_s\bar D_s$ and $D_s\bar D_s\rightarrow D_s\bar D_s$ amplitudes
immediately above the $D_s\bar D_s$ threshold. Resonance parameters of this pole share 
similar features to that of $\chi_{c0}(3930)$/X(3915), both of which are below 
the $D_s\bar D_s$ threshold in the physical world. Most importantly, these states have 
narrow widths and were not observed in the $D\bar D$ final states, resulting from their 
large interaction with the $D_s\bar D_s$ channel.

In the right side of Figure \ref{rqcdscalar1}, we present the extracted charmonium spectrum 
in the scalar channel together with the experimental observations. Several simplifying 
assumptions were made in this study of the excited charmonium spectrum, including 
heavier than physical light and charm quarks, lighter than physical strange quarks, 
and neglecting scattering channels ($J/\psi\omega$ and $\eta_c\eta$). Furthermore, the 
scattering amplitudes were extracted with simple parametrizations, and the model 
dependence of the results was not investigated in detail. It would be exciting to see 
how these findings evolve when various approximations are relaxed in future lattice simulations.

\section{Summary}\label{summar}

Precision measurements of ground-state hadrons are now well established using lattice QCD 
methodology. We summarized various lattice results for ground-state charmed and bottom baryons. 
Recent highlights has been the $\Xi_{cc}$ baryon and the excited $\Omega_{c}$ baryons discussed 
in Section \ref{cbbaryons}. Lattice predictions for the mass of $\Xi_{cc}$ baryon (and several other 
baryons) agree very well with the experimental value. We also discuss some of the recent lattice 
efforts in the heavy tetraquark and dibaryon sectors in Section \ref{beyond}. Various lattice studies 
points to the existence of strong interaction stable doubly bottom tetraquark system. More 
calculations are anticipated with regard to the doubly charm and charmed-bottom four quark systems.  

In Section \ref{mesres}, we discuss the status of lattice investigations of charmonium excited states with 
special focus on the vector and scalar channels. While the studies of the vector channel serves 
as a proof of principle, the studies of scalar charmonium spectra are aimed at possible explanations 
of the low-lying scalar charmonium-like exotic candidates within the framework of QCD. The highlight in 
these studies is the observation of a shallow $D\bar D$ bound state and two other pole singularities 
possibly related to X(3860) and $\chi_{c0}(3930)$/X(3915).

\acknowledgments
I thank Nilmani Mathur and Daniel Mohler for their careful reading of the manuscript. 

\bibliographystyle{JHEP}
\bibliography{references}

\end{document}